\documentclass{rmaa}

\title{Proper Motions of the Ansae in the 
Planetary Nebula NGC~7009}

 \author{Luis F. Rodr\'\i guez and Yolanda G\'omez
  \affil{Centro de Radioastronom\'{\i}a y Astrof\'{\i}sica, UNAM, Morelia}
 }

\fulladdresses{
\item Yolanda G\'omez and Luis F. Rodr\'\i guez: Centro de Radiostronom\'{\i}a
y Astrof\'\i sica, 
UNAM, A. P. 3-72, 
(Xangari), 58089 Morelia, Michoac\'an, M\'exico
(y.gomez; l.rodriguez@astrosmo.unam.mx).
}

\shortauthor{Rodr\'iguez \& G\'omez}
\shorttitle{Proper Motions of Ansae in NGC 7009}

\SetVolume{43} \SetFirstPage{001} \SetYear{2007}
\ReceivedDate{2006 August 04} 
\AcceptedDate{2006 November 23}

\resumen{Para la nebulosa planetaria NGC 7009, 
presentamos una comparaci\'on de dos conjuntos de datos sin publicar del archivo
del Very Large Array tomados con una separaci\'on temporal de
8.09 a\~nos para confirmar los movimientos propios observados en
el \'optico en sus ansas. Determinamos valores de
23$\pm$6 y 34$\pm$10 mas a\~no$^{-1}$ para las ansas este y oeste, respectivamente.
Hay evidencia marginal de que
las densidades de flujo de los chorros que conectan a las ansas con
el cuerpo principal de la nebulosa disminuyeron en un
30\% en el periodo que separa a las dos observaciones.
Tambien establecemos un l\'\i mite superior a la expansi\'on del cuerpo principal de la
nebulosa planetaria, obteniendo un l\'\i mite inferior de
$\sim$700 pc para la distancia a ella. 
}
 
\abstract{For the planetary nebula NGC 7009,
we present a comparison of two unpublished
Very Large Array archive
data sets taken with a time separation of 8.09 years to confirm
the proper motions of its ansae 
observed in the optical. We determine values of
23$\pm$6 and 34$\pm$10 mas yr$^{-1}$ for the
eastern and western ansae, respectively.
There is marginal evidence suggesting that the flux densities of the jets that connect
the ansae with the main body of the nebula diminished
in about 30\% over the period between the two observations.
We also set an upper limit to the expansion of the main
body of the planetary nebula, setting a lower limit of $\sim$700 pc 
for its distance.
}
      

\begin{document}

\maketitle

\section{Introduction}

NGC 7009 (PN G037.7-34.5), the "Saturn Nebula", is a well-studied elliptical PN, which has 
a jetlike system as well as a pair of low-ionization knots along its major axis. 
This type of knots, also known as ansae (Aller 1941), are now known to be
present in many planetary nebulae and have received the names of
bipolar, rotating, episodic jets
BRETS (L\'opez, V\'azquez, \& Rodr\'\i guez 1995; L\'opez 1997), and 
fast, low ionization emission regions 
(FLIERS; Balick et al. 1993; 1994). There are several models proposed to
explain the origin of these structures, but no one is generally
accepted (for a discussion, see Steffen, L\'opez, \& Lim 2001).
The radial velocity observations of several of these objects by Balick, Preston, \&
Icke (1987) indicated velocities in the order of several tens of km s$^{-1}$
with respect to the systemic velocity of the planetary nebula.
On the other hand,
proper motions of these knots have been measured only for a handful of objects
(KjPn 8; Meaburn 1997, Hen 2-90; Sahai et al. 2002, NGC 7009;
Fern\'andez, Monteiro, \& Schwarz 2004). In particular, at
radio wavelengths there are, to our knowlewdge, no measurements of
proper motions of ansae in planetary nebulae. In this paper we present
such measurement for the ansae in NGC 7009. We also discuss an apparent change
in the flux density of the jets of this object and study the expansion of the
main body of the nebula, setting a lower limit to its distance. 
 
\section{Observations}

The two sets of observations used in this study
were taken from the archive
of the VLA of the NRAO\footnote{The National Radio 
Astronomy Observatory is operated by Associated Universities 
Inc. under cooperative agreement with the National Science Foundation.}.
The epochs of the observations were 
1989 March 28 (epoch 1989.24) and 1997 April 29
(epoch 1997.33), with a time separation of 8.09 years. 
Both sets of observations were made at 3.6 cm in the B configuration
and each has a total of $\sim$7 hours of on-source integration.
In both epochs the source 1331+305 was used as an absolute amplitude
calibrator (with an adopted flux density of 5.21 Jy)
and the source 2131-121    
was used as phase calibrator
(with a bootstrapped flux density of 
3.19$\pm$0.01 Jy for the first epoch and of
2.68$\pm$0.02 Jy for the second epoch).

The data were reduced using the standard VLA procedures in
the software package Astronomical Imaging Processing System
(AIPS) of NRAO and then cross-calibrated
using the procedure of Masson (1986; 1989a; 1989b).

\begin{figure*}
\centering
\includegraphics[scale=0.55, angle=0]{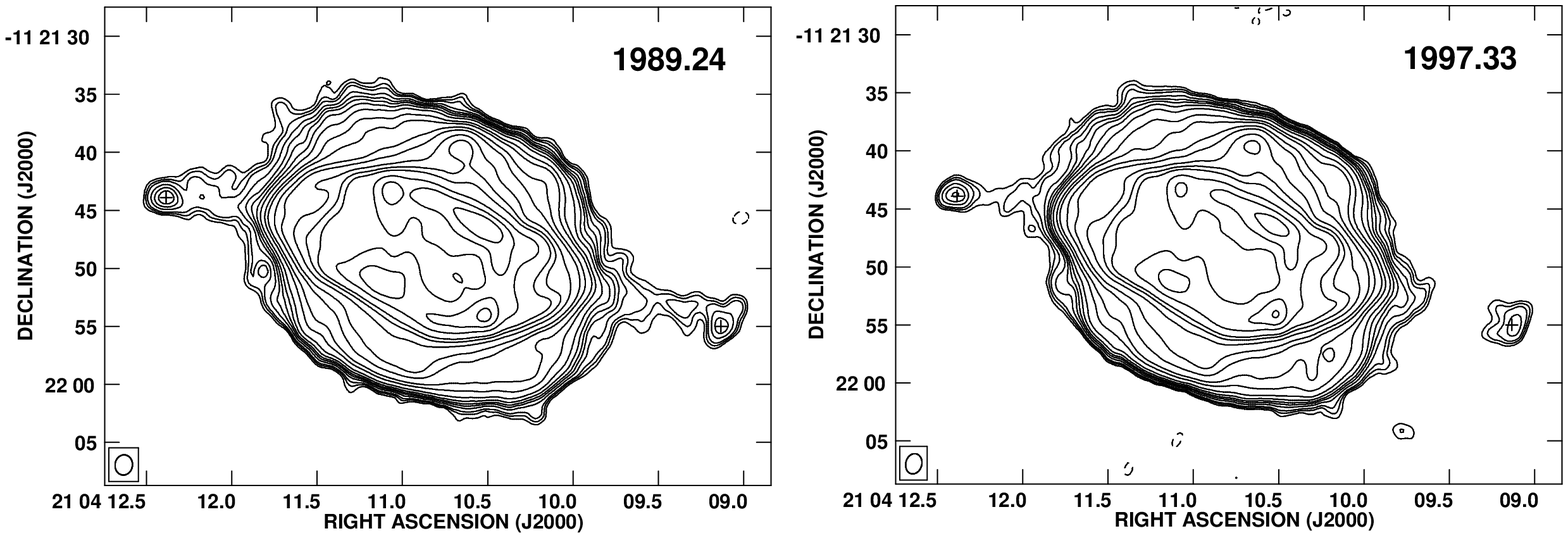}
\includegraphics[scale=0.55, angle=0]{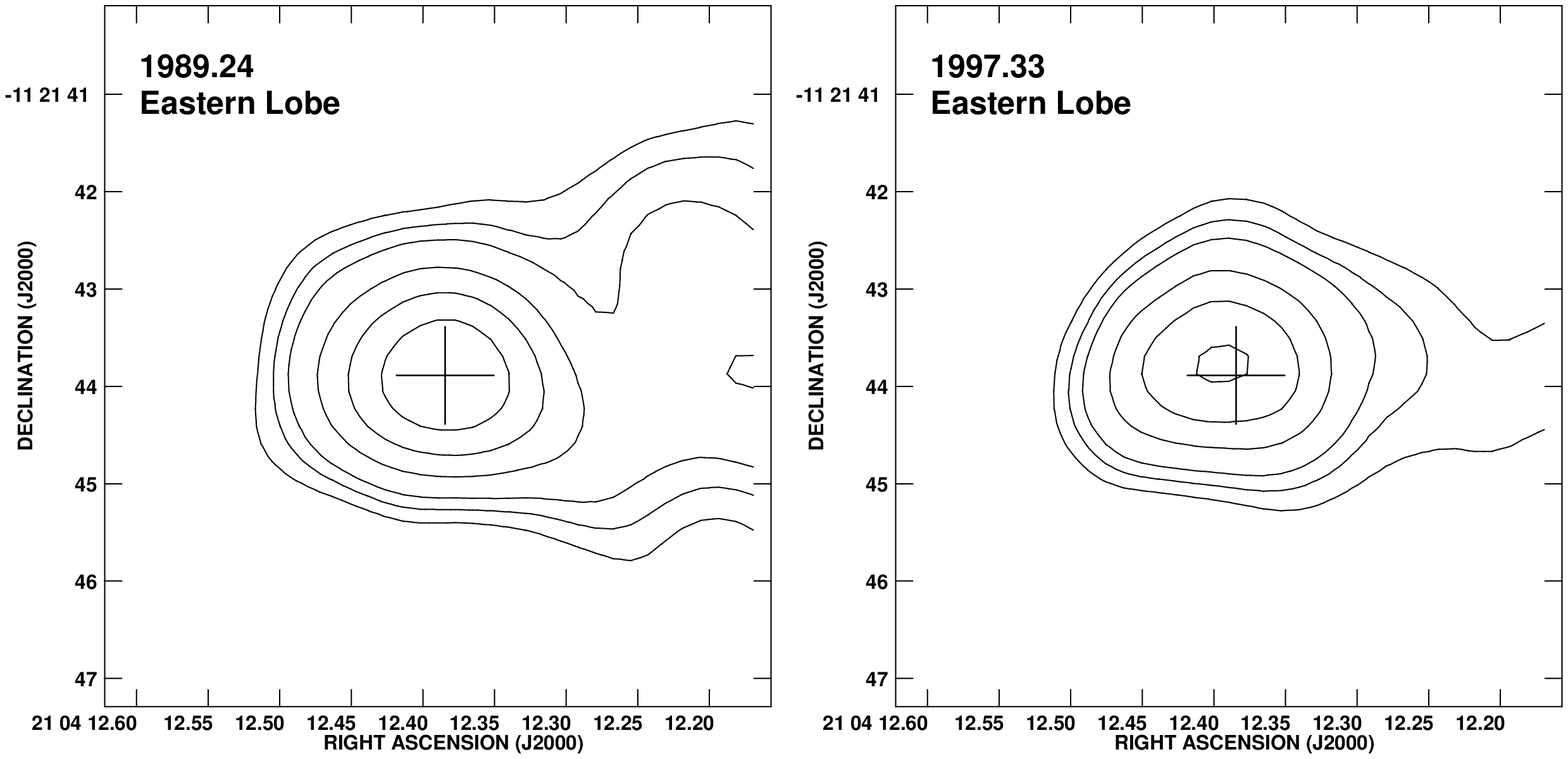}
\includegraphics[scale=0.55, angle=0]{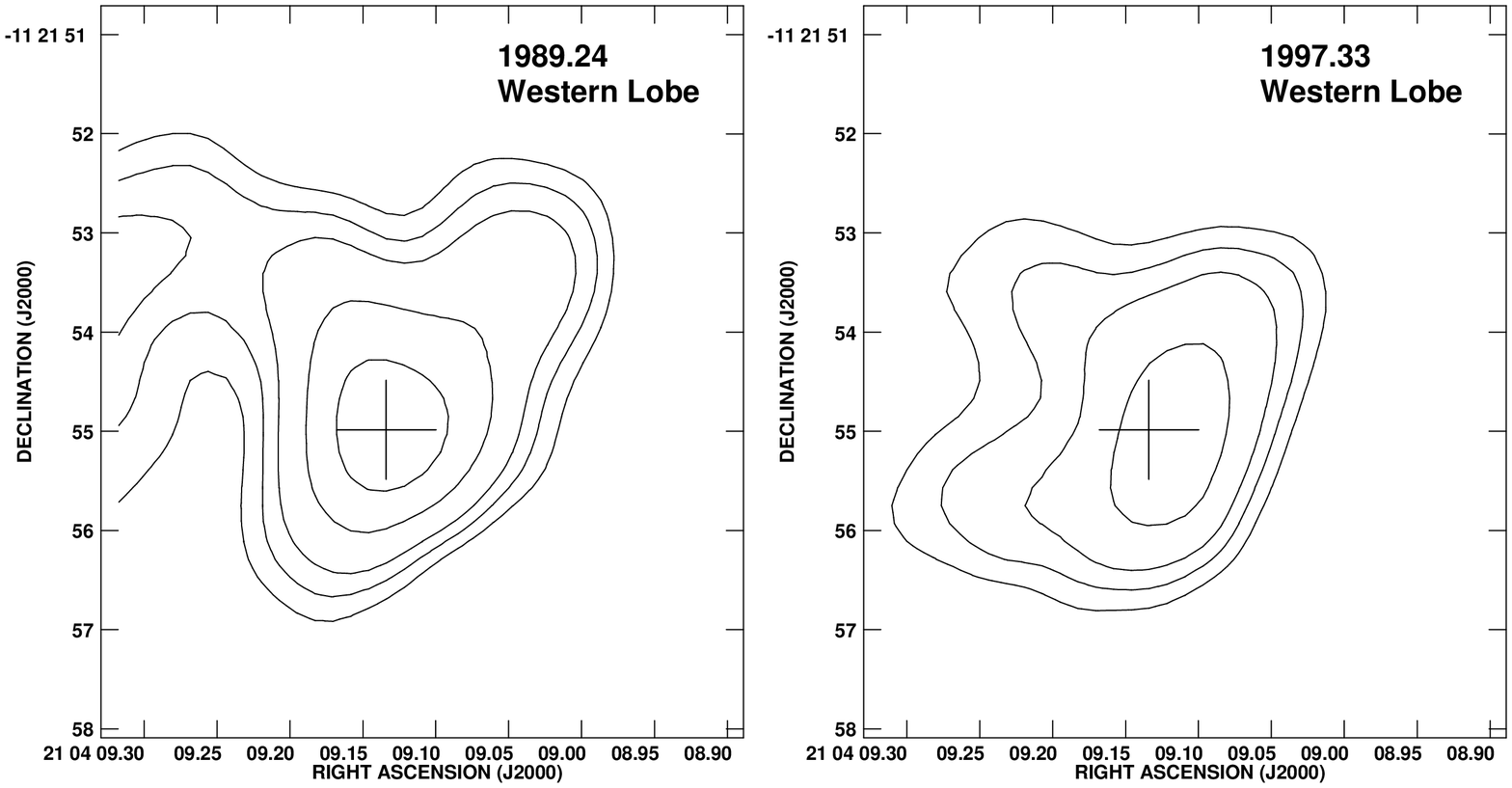}
 \caption{3.6 cm contour images of NGC 7009 for 1989.24 (left)
and 1997.33 (right). The top part of the panel shows the whole of the nebula,
while the middle and bottom parts of the panel show the eastern and western 
lobes, respectively. The crosses mark the peak position
of the lobes for 1989.24. Note the small displacement of the peak position
of the lobes for 1997.33. The contours are -4, 4, 5,
6, 8, 10, 12, 15, 20, 30, 40,
60, 80, 120, 200, 300, 400, and 500
times 19.5 $\mu$Jy beam$^{-1}$, the average rms noise of the two images.
The synthesized beams ($1\rlap.{''}74 \times 1\rlap.{''}50$
with a position angle of $-8^\circ$ for 1989.24 and
$1\rlap.{''}76 \times 1\rlap.{''}38$
with a position angle of $-15^\circ$ for 1997.33) are shown in the bottom left corner
of the images at the top of the panel.
}
  \label{fig1}
\end{figure*}

\begin{table*}[htbp]
  \setlength{\tabnotewidth}{0.9\textwidth} 
  \tablecols{6} 
  \caption{Proper Motions\lowercase{$^a$} of the Ansae}
  \begin{center}
    \begin{tabular}{lccccc}\hline\hline
Ansa & $\Delta_{RA}$({''}) & $\Delta_{DEC}$({''}) & $\Delta_{TOTAL}$({''}) 
& P.A. ($^\circ$) & $\mu$ (mas yr$^{-1}$) \\
\hline
Eastern &  +0.14$\pm$0.05 & +0.11$\pm$0.05 & 0.18$\pm$0.05 & 52$\pm$16 & 23$\pm$6 \\
Western &  -0.23$\pm$0.08 & -0.14$\pm$0.08 & 0.27$\pm$0.08 & 239$\pm$17 & 34$\pm$10 \\
\hline\hline
\tabnotetext{a}{$\Delta_{RA}$ is the displacement in right ascension,
$\Delta_{DEC}$ is the displacement in declination,
$\Delta_{TOTAL}$ is the total displacement, P.A. is the 
position angle of the displacement, and $\mu$ is the total proper
motion.}
    \label{tab:1}
    \end{tabular}
  \end{center}
\end{table*}
\section{Interpretation and Results}

\subsection{Proper Motions of the Ansae}

In Figure 1 we present 3.6 cm images for the two epochs
analyzed. Three main structures are evident in the images.
The first is the bright, main body of the nebula with an elliptical shape
and angular dimensions of about $32{''} \times 24{''}$. 
The second structure is constituted by the faint jets that 
emanate from the main body of the nebula and extend about
$8{''}$ to the east and west. Finally, the third structure are the
ansae, with angular dimensions of a few arc sec and that possibly are
the termination points of the jets.

The images were made with the ROBUST weighting parameter
(Briggs 1995) of AIPS
set to 0 and a tapering of 150 k$\lambda$ in the \sl (u,v) \rm plane to smooth
the angular resolution and enhance the signal-to-noise
ratio of the relatively faint and extended ansae.
The expanding proper motions of the ansae are evident in the images.
In Table 1 we present the values of these proper motions. The peak
positions of the ansae have been obtained with the task MAXFIT
of the AIPS software package. The errors in the positions were
estimated following Condon (1997) and Condon \& Yin (2001). There are
two previous reports of these proper motions in the literature.
Liller (1965) estimated a value of $\sim$16 mas yr$^{-1}$ for the proper motion
of the ansae in NGC 7009. A more modern determination is that obtained by
Fern\'andez et al. (2004), who used HST images to obtain a value
of 28$\pm$8 mas yr$^{-1}$ for the eastern ansa (the western ansa was
not included in one of the archive images they used). This value compares well 
with the value of 23$\pm$6 mas yr$^{-1}$ derived by us (see Table 1).
Assuming a distance of 0.86$\pm$0.34 kpc, the average of 14 available values
(Acker et al. 1992; Fern\'andez et al. 2004), and taking into account
the error from our proper motions as well as
the relatively large error in the average distance, we obtain 
crude estimates for the velocities in the plane of the sky of 100$\pm$50 and
140$\pm$70 km s$^{-1}$, 
for the east and west ansae, respectively.

If we assume that the ansae move ballistically (i. e. with no acceleration or
deceleration) we estimate an age of $\sim$850 years
if they originated from the central star and of $\sim$300 years if they originated from
the edge of the main body of the planetary nebula.

\subsection{A Decrease in the Emission from the Jets?}

There is marginal evidence in Figure 1 that, between the two epochs
observed, the emission from the 
jets that connect the ansae with the main body of the
planetary nebula has decreased. For both the east and west jets, we find that the 
flux density decreased
from $\sim$0.9$\pm$0.1 mJy in 1989.24 to $\sim$0.6$\pm$0.1 mJy in 1997.33, a decrease
of about 30\%. 
The ansae themselves, at the tip of the jets, do not show
evidence of variation within the noise.

The possible decreases in the flux densities from the jets
over the 8-year period
are marginally consistent with recombination theory.
The electron densities of the jets have been estimated to
be between 1,000 and 4,000 cm$^{-3}$
(Bohigas, L\'opez, \& Aguilar 1994; Balick et al. 1994; Gon\c{c}alves et al.
2003; Sabbadin et al. 2004), implying recombination
timescales from 30 to 120 years.
Evidence of variation in the surroundings of NGC 7009 has been found by
Bohigas et al. (1994), who report a previously undetected [SII] condensation about
20{''} to the NE
of the main body of the nebula.  
On the other hand, as a whole, the planetary nebula
has a flux density of 654$\pm$3 mJy in 1989.24 and of
649$\pm$5 mJy in 1997.33, so the total flux density remains constant
within $\leq$1 \%.  

\subsection{An Upper Limit to the Expansion of the Nebula}

\begin{figure*}
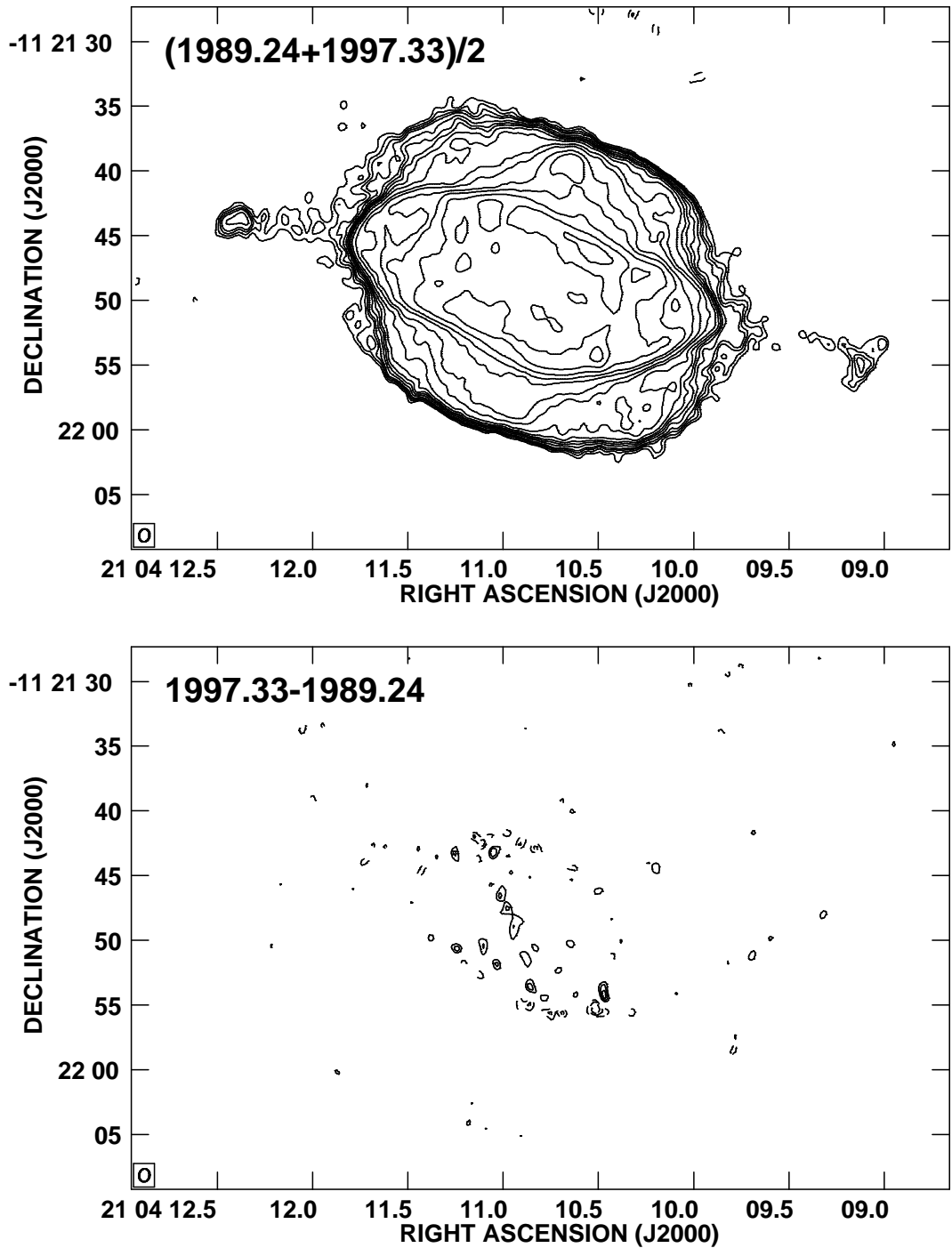

\centering
\includegraphics[scale=0.80, angle=0]{N7009AVE.PS}
\includegraphics[scale=0.80, angle=0]{N7009DIF.PS}
 \caption{3.6 cm contour images of NGC 7009 for all the
data (top) and the difference 1997.33$-$1989.24 (bottom).
The contours are  -4, -3,
3, 4, 5, 6, 8, 10, 12, 15,
20, 30, 40, 60,
80, 120, 200, and 300
times 12 $\mu$Jy beam$^{-1}$ for the top image
and -5, -4, -3, 3, 4, and 5
times 20 $\mu$Jy beam$^{-1}$ for the bottom image.
The synthesized beam ($1\rlap.{''}00 \times 0\rlap.{''}81$
with a position angle of $-10^\circ$)
is shown in the bottom left corner
of the images.
}
  \label{fig2}
\end{figure*}

Comparison of data taken with years or decades of separation
often reveal the expansion signature in planetary nebula
(e. g. Guzm\'an, G\'omez, \& Rodr\'\i guez 2006). When subtracting
the older data from the more recent one, the characteristic
expansion signature shows as an outer region of positive
emission, surrounding an inner region of ``negative'' emission.
In Figure 2 we show an average image from the data of the two
epochs (obtained by concatenating both data sets) and a 
difference image, 1997.33$-$1989.24 (obtained by subtracting
the clean components of the 1997.33 image from the \sl(u,v) \rm
data of 1989.24). These images were made with the ROBUST weighting parameter
of AIPS
set to 0 and no tapering. Although there is evidence of residual
emission in the brightest parts of the nebula, we failed to detect
a clear expansion signature (see bottom of Fig. 2). An upper 
limit to the expansion can be
obtained following G\'omez, 
Rodr\'\i guez, \& Moran (1993) and Guzm\'an et al. (2006). 
We assume that the nebula expands in a self-similar way by a factor of
(1 + $\epsilon$), where $\epsilon << 1$.
From the non detection of an expansion pattern, we estimate
that $\epsilon \leq 0.007$. A larger value of $\epsilon$ would produce
residuals in the difference image larger than those observed.
A lower limit to the distance, $D$, of the planetary nebula is then given by

$$\bigg[\frac {D} {pc}\bigg] \geq 0.211\bigg[\frac {\theta} 
{''}\bigg]^{-1} \bigg[\frac {v_{exp}} 
{km~s^{-1}}\bigg] \bigg[ \frac {\Delta t} {yr} \bigg] \epsilon^{-1},$$

\noindent where $\theta$ is the
average angular radius of maximum emission,
$v_{exp}$ is the expansion velocity of the nebula at the radius of
maximum emission, and $\Delta t$ is the time interval between observations.
NGC 7009 is an elliptical nebula and we approximate the angular radius of
maximum emission as the geometric mean of the major and minor semiaxes.
From Figure 2, we estimate $\theta \simeq 12{''}$. From Weedman
(1968) we find that at this angular distance 
$v_{exp} \simeq$ 35 $km~s^{-1}$. We then obtain $D \geq$ 700 pc,
a lower limit consistent with the weighted average of 14 available values
(Acker et al. 1992; Fern\'andez et al. 2004), 860$\pm$340 pc. 
It should be noted that the expansion parallax distance technique can
be affected by the motion of the shock or the ionization fronts
(Mellema 2004) and that our lower limit could be underestimated by 20-30\%. 

\section{Conclusions}

We analyzed archive VLA observations of the 
planetary nebula NGC 7009 for two epochs separated
by 8.09 years.
Our main conclusions are: 

1. We measured the proper motions of the ansae, obtaining
values of 23$\pm$6 and 34$\pm$10 mas yr$^{-1}$ for the
eastern and western ansae, respectively. This is the first time
that proper motions of ansae in planetary nebulae
are determined with radio observations. 

2. There is marginal evidence suggesting that the flux densities of the jets that  
connect
the ansae with the main body of the nebula diminished
in about 30\% over the 8.09 years spanned by the observations.

3. We failed to detect the expansion of the main body
of the planetary nebula, obtaining a lower limit of $\sim$700 pc
for its distance.


\acknowledgments
We thank an anonymous referee for the careful reading of the manuscript.
LFR and YG acknowledge the support
of DGAPA, UNAM, and of CONACyT (M\'exico).
This research has made use of the SIMBAD database, 
operated at CDS, Strasbourg, France.



\begin{thebibliography}

\bibitem{ac92} Acker, A., Marcout, J., Ochsenbein, F., Stenholm, B., \& Tylenda, R. 
1992, Strasbourg-ESO Catalogue of Galactic Planetary Nebulae, ESO, Garching

\bibitem{al41} Aller, L. H. 1941, ApJ, 93, 236

\bibitem{ba87} Balick, B., Preston, H. L., \& Icke, V. 1987, AJ, 94, 1641

\bibitem{ba93} Balick, B., Rugers M., Terzian Y., \& Chengalur J. N., 1993, ApJ, 411, 778

\bibitem{ba94} Balick, B., Perinotto, M., Maccioni, A., Terzian, Y., \& Hajian, A.  
1994, ApJ, 424, 800

\bibitem{be01} Beltr\'an, M. T., Estalella, R., Anglada, G., Rodr\'\i guez, L. F., \& 
Torrelles, J. M. 2001, AJ, 121, 1556

\bibitem{bo94} Bohigas, J., L\'opez, J. A., \& Aguilar, L. 1994, A\&A, 291, 595

\bibitem{br95} Briggs, D. 1995, Ph.D. thesis, New Mexico Inst. of 
Mining and Technology

\bibitem{co97} Condon, J. J. 1997, PASP, 109, 166

\bibitem{co01} Condon, J. J. \& Yin, Q. F. 2001, PASP, 113, 362

\bibitem{fe04} Fern\'andez, R., Monteiro, H., \& Schwarz, H. E. 2004,
ApJ, 603, 595

\bibitem{grm93} G\'omez, Y., Rodr\'\i guez, L. F., \& Moran, J. M. 1993, ApJ, 416, 620

\bibitem{go03} Gon\c{c}alves, D. R., Corradi, R. L. M., Mampaso, A.,
\& Perinotto, M. 2003, ApJ, 597, 975

\bibitem{gu06} Guzm\'an, L., G\'omez, Y., \& Rodr\'\i guez, L. F. 2006,
RevMexA\&A, 42, 127

\bibitem{li65} Liller, W. 1965, PASP, 77, 25

\bibitem{lo95} L\'opez J. A., V\'azquez R., \& Rodr\'\i guez L. F., 1995, ApJ, 455, L63

\bibitem{lo97} L\'opez, J. A. 1997, in IAU Symp. 180, Planetary Nebulae, 
ed. H. J. Habing \& H. J. G. L. M. Lamers (Dordrecht: Kluwer), 197

\bibitem{ma86}  Masson, C. R. 1986, ApJ, 302, L27

\bibitem{ma89a} Masson, C. R. 1989a, ApJ, 336, 294

\bibitem{ma89b} Masson, C. R. 1989b, ApJ, 346, 243

\bibitem{me97} Meaburn, J. 1997, MNRAS, 292, L11

\bibitem{mel04} Mellema, G. 2004, A\&A, 416, 623 

\bibitem{sa04} Sabbadin, F., Turatto, M., Cappellaro, E., Benetti, S., \& Ragazzoni, R.
2004, A\&A, 416, 955

\bibitem{sa02} Sahai, R., Brillant, S., Livio, M., Grebel, E. K., 
Brandner, W., Tingay, S., \& Nyman, L.-\AA. 2002, ApJ, 573, L123

\bibitem{st01} Steffen, W., L\'opez, J. A., \& Lim, A. 2001, ApJ, 556, 823 

\bibitem{we68}  Weedman, D. W. 1968, ApJ, 153, 49


\end{thebibliography}
\end{document}